# Salt-assisted vapor-liquid-solid growth of one-dimensional van der Waals materials

Thang Pham[1*], Kate Reidy[1], Joachim D. Thomsen[1], Baoming Wang[1], Nishant Deshmukh[2], Michael A. Filler[2*] and Frances M. Ross[1*]

1. Department of Materials Science and Engineering, Massachusetts Institute of Technology, Cambridge, MA 02139, USA
2. School of Chemical & Biomolecular Engineering, Georgia Institute of Technology, Atlanta, GA 30332, USA

Corresponding authors. Email: thangpt88@gmail.com (T.P.), michael.filler@chbe.gatech.edu (M.A.F.), fmross@mit.edu (F.M.R.)

**Abstract**

We have combined the benefits of two catalytic growth phenomena to form nanostructures of transition metal trichalcogenides (TMTs), materials that are challenging to grow in a nanostructured form by conventional techniques, as required to exploit their exotic physics. Our growth strategy combines the benefits of vapor-liquid-solid (VLS) growth in controlling dimension and growth location, and salt-assisted growth for fast growth at moderate temperatures. This salt-assisted VLS growth is enabled through use of a catalyst that includes Au and an alkali metal halide. We demonstrate high yields of $NbS_3$ 1D nanostructures with sub-ten nanometer diameter, tens of micrometers length, and distinct 1D morphologies consisting of nanowires and nanoribbons with [010] and [100] growth orientations, respectively. We present strategies to control the growth location, size, and morphology. We extend the growth method to synthesize other TMTs, $NbSe_3$ and $TiS_3$, as nanowires. Finally, we discuss the growth mechanism based on the relationships we measure between the materials characteristics (growth orientation, morphology and dimensions) and the growth conditions (catalyst volume and growth time). Our study introduces opportunities to expand the library of emerging 1D vdW materials and their heterostructures with controllable nanoscale dimensions.

## INTRODUCTION

The discovery of graphene has stimulated interested in van der Waals (vdW) bonded materials that can be mechanically or chemically exfoliated to obtain single layer sheets (*1, 2*). Beyond well-known examples, such as two-dimensional (2D) transition metal dichalcogenides (TMDs), are the less-studied vdW materials based on one-dimensional (1D) motifs that show exotic properties and promising applications owing to their intrinsic anisotropic character (*3–5*). The prototypical family of these so-called "1D vdW" materials is the transition metal trichalcogenides (TMTs) with the formula $MX_3$, such as $NbSe_3$, $TiS_3$ and $HfTe_3$ (*4, 6*). These materials contain covalently bonded triagonal prismatic chains of metal (M) and chalcogens (X) further assembled, through weaker inter-chain covalent bonds, into bilayer sheets that are stacked to result in a mixture of 2D and 1D characteristics (*7*). Even in the bulk form, many of these crystals exhibit intriguing electronic, optical, and thermal properties uniquely derived from their 1D character (*3–5*). This includes hosting charge and spin density waves(*4*), superconductivity (*4*) and spatially confined phonon density of states(*3*), to name just a few.

Moreover, nanoscale, exfoliated forms of several TMTs have shown exotic optoelectronic (*8*, *9*) and thermal transport characteristics (*10*), for instance superdiffusive thermal transport in $NbSe_3$ nanowires with diameter smaller than 20 nm (*11*), highlighting the importance of nanoscale confinement in these materials (*3*, *7*, *12*, *13*).

$NbS_3$ is an interesting TMT due to its extensive library of polymorphs and structure-dependent properties (*14–16*). Five distinct phases of $NbS_3$ have been observed experimentally and many more stable phases were recently predicted (*15*). Different polymorphs usually have different degrees of metal pairing (Nb-Nb distance) along the chain, or different bilayer stacking order. These structural differences result in a wealth of electronic properties, for example three distinct charge density wave instabilities in type-II $NbS_3$, a metallic phase with basic structure belonging to the space group P2$_1$/$m$ (*14*, *17*). Despite these fascinating structures and properties, $NbS_3$ is one of the least reported members in the family of TMTs in terms of both fundamental studies and potential applications, in part because of the lack of a synthesis method that enables precise manufacturing of the material (*18*). Successful integration of $NbS_3$ into next-generation devices that make use of its nanoscale properties would benefit from a deterministic synthesis method that allows control over the material's structural phase, chemistry, size, and placement (*3*).

Bulk TMTs have mainly been synthesized by chemical vapor transport and flux growth(*5*, *6*). While these methods produce single crystals of high purity, they face several issues in scalable manufacturing of nanostructures (*3*, *5*). Long times of days to weeks are required to grow the crystals, and additional steps such as exfoliation (mechanically or chemically) and transfer are needed to place nanoscale samples onto target substrates for further characterization and applications. The need for post-growth processing often results in a lack of control of the resulting nanomaterial in terms of its size and interface quality, hindering integration with other material systems (*5*, *19*).

A successful strategy to create 1D nanostructures of semiconductors with controlled sizes and placement on a substrate has been the catalytic process of vapor-liquid-solid (VLS) growth (*20*). The working principle of VLS depends, at its most general, on the formation of a liquid eutectic between the growth material and a catalyst patterned onto the substrate at the desired location of the nanowire (*20*). With Au as a common catalyst, growth materials have ranged from group IV and III-IV semiconductors (*21*, *22*) such as Si, Ge, and GaAs, usually supplied by chemical vapor deposition (CVD) precursors, to binary oxides, e.g., ZnO (*23*, *24*). A versatile feature of VLS is the ability to create heterostructures; even chemically dissimilar materials have been interfaced (*25–27*). For layered materials, the VLS mechanism has been demonstrated for synthesis of nanowires containing low-melting metals, such as Ga (GaS (*28*), GaSe (*29*)), In ($In_2Se_3$ (*30*)), Bi ($Bi_2Se_3$ (*31*)) and Ge (GeS (*32*, *33*), GeSe (*34*)). However, for synthesis of TMTs the technique is more challenging, since the metals of interest have very high melting points and low vapor pressure (e.g., Nb and Ti) and do not form liquid eutectic compositions with a metal such as Au in the typical CVD temperature range of 600-900°C. As such, to the best of our knowledge, there has been no study of TMT nanostructures synthesized by VLS or any other method that can provide a high degree of control over the material's structure, chemistry, and dimensions.

Here we report a modification of the VLS mechanism that enables synthesis of transition metal trichalcogenides, exemplified by $NbS_3$ 1D nanostructures, developed by adapting a strategy of salt-assisted growth that has been demonstrated for other materials, specifically TMDs. The technique is based on the use of VLS catalysts with conventional solid sources for Nb and S that are combined with an alkali metal halide, NaCl. As discussed below, the salt is known to react with transition metal/metal oxides (Nb/NbO$_x$) to form intermediate products that have lower melting points (*35, 36*). This improves the vapor pressure and increases the mass flow of the metal (Nb) precursor, resulting in growth over short times (time scales of minutes) at temperatures (650-750°C) much lower than would be required in the absence of the salt. Under these conditions, the addition of a metallic catalyst enables VLS growth with control of 1D nanostructure sizes and growth locations.

We show that this hybrid growth strategy, which we refer to as *salt-assisted VLS growth*, results in high-density 1D materials with sub-ten nanometer diameter and lengths up to several tens of micrometers, the regime in which strongly correlated physics starts to emerge (*3*). Additionally, using electron microscopy techniques (electron diffraction and atomic-resolution scanning transmission electron microscopy) we show that the materials exhibit several distinct 1D morphologies, such as wires and ribbons, and two growth orientations, namely [010] and [100] directions. We present strategies to control the growth of 1D NbS$_3$, varying the nanostructure size through the growth time, the 1D morphology through the catalyst size, the growth location via catalyst patterning, and we discuss the effect on growth of the substrate by using both bulk substrate (SiO$_x$/Si, SiN, sapphire) and 2D substrate (few-layer graphene and hexagonal boron nitride). We demonstrate the synthesis of other TMT nanowires, namely NbSe$_3$ and TiS$_3$, using the same approach. Finally, we propose a growth mechanism consistent with these findings based on the relationships we measure between the NbS$_3$ growth orientation, morphologies, catalyst diameter, growth time, and nanowire dimensions. The dimensions we obtain are within the range needed to display the materials' unusual electron, phonon, and spin transport properties (*37*). Our study introduces an opportunity to expand the library of vdW materials towards emerging 1D materials, and we anticipate that the deterministic synthesis will enable the integration of 1D quantum materials into next generation quantum electronics. We also propose this as a more general approach for exploring nanowire synthesis of group IV and V heavy metal compounds (e.g., oxides of Mo, W, Ta, Hf) based on an alloy catalyst containing both a metal and an alkali metal halide.

## RESULTS
### Structures and growth orientations

We show in Figure 1 an overview of the types of structures that we have obtained through salt-assisted VLS growth of NbS$_3$. Figure 1A illustrates growth over a large area, showing a high density of 1D structures with spherical particles at one end, indicative of the VLS mechanism (*22*). In this example, diameters in the range of 10-500 nm and lengths up to tens of micrometers were obtained for 30 minutes growth at 725°C using 10-nm Au catalyst film. Figure S1 shows the selectivity of this growth, with locations controlled through patterning of Au on a SiO$_x$/Si substrate.

This growth and all others presented below were obtained in the deposition system shown in Fig. S1A, with a process that is further described in the Methods. We studied growth on both bulk substrates (amorphous SiO$_x$/Si, SiN, sapphire, and mica) and on 2D vdW materials (single to few-layer graphene and hexagonal BN) transferred onto SiN TEM grids (Fig. S1B). Au with thickness in the range 0-50 nm is evaporated onto these substrates to provide the VLS catalyst metal; on heating, these films agglomerate into droplets from which the nanostructures grow. The solid source for Nb is mixed with NaCl to act as a growth promoter. The salt is intended to react with the Nb powder precursor to form intermediate products with lower melting point, aiming to enhance the vapor pressure and increase the mass flow (*38*, *39*). Typically, 50 mg of Nb powder is mixed with 25 mg of NaCl, and the mixture is loaded into an alumina boat. A growth substrate is placed upside down directly above the precursor mixture and the Nb/NaCl/substrate boat is set to a temperature of 650-800°C. 150-200 mg of S crystals are placed upstream and heated to 130-150°C. 100-150 sccm of high purity Ar is used as the carrier gas for growth times varying from 5 to 45 minutes. We note that below 650°C, the evaporation rate is very limited and the growth results in short and small structures. On the other hand, above 800°C, we observe mostly $NbS_2$ phases (Fig. S2), in agreement with a recent report showing the conversion of $NbS_2$ phase from $NbS_3$ at high temperature (1000°C) (*40*). We therefore find an optimum temperature of 725°C and this was used for all of the growths discussed below, unless indicated otherwise.

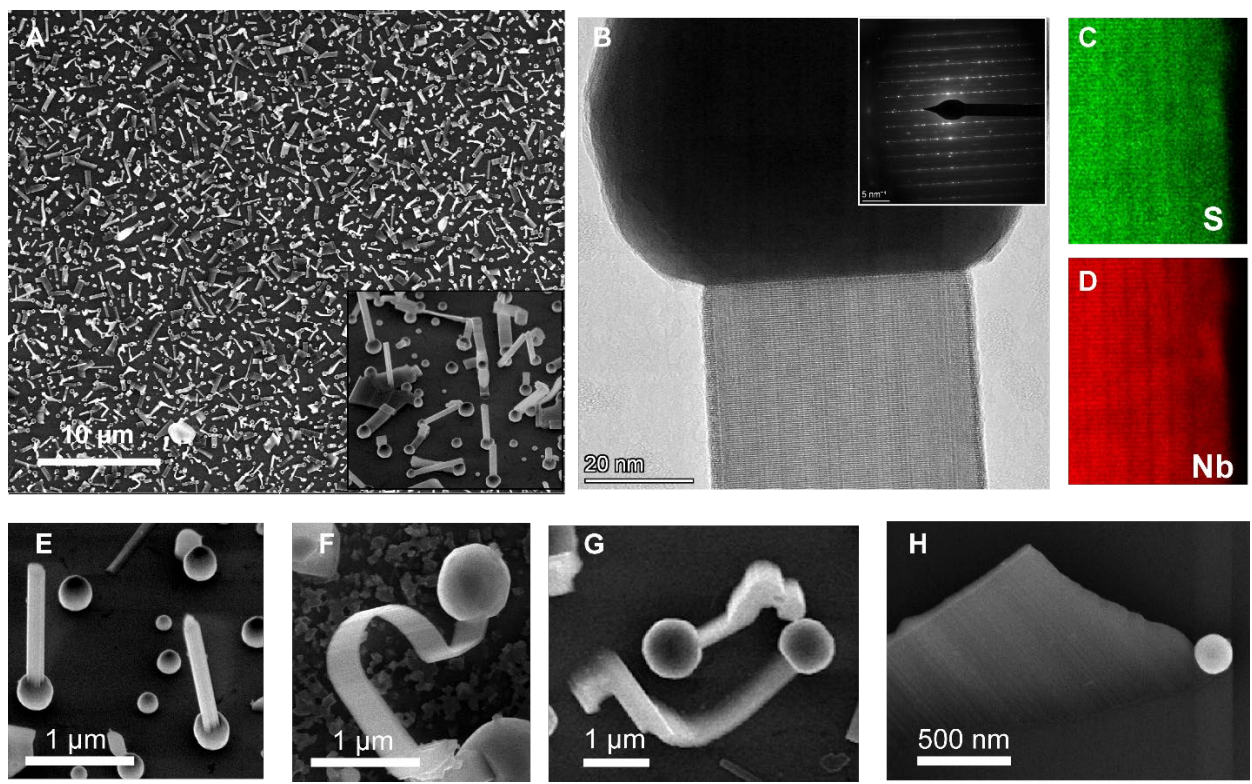

**Fig. 1. VLS growth of $NbS_3$ one-dimensional materials.** (A) A representative SEM image of 30-min growth on a $SiO_x$/Si substrate showing high density of 1D nanomaterials. Inset: High magnification SEM image displays the presence of nanoparticles residing at one end of the 1D materials. (B) High-resolution TEM image with electron diffraction pattern inset. (C-D) Atomic-resolution EDS chemical composition mapping of a $NbS_3$ nanowire. (E-H) Typical morphologies of as-synthesized $NbS_3$ 1D nanostructures: (E) nanowires having growth orientation [010], and (F-H) nano-ribbons having growth orientation [100].

Atomic structure and chemistry of the as-synthesized materials were measured by electron microscopy. Atomic-resolution TEM images and electron diffraction (Fig. 1B) reveal the Nb-Nb repeat distance along the chain direction is b= 3.36 ± 0.12 nm, while the other lattice parameters are a = 9.63 ± 0.21 nm and c = 19.62 ± 0.29 nm. These parameters indicate that the deposited material is monoclinic $NbS_3$ ($NbS_3$-II phase), a stable and metallic phase of $NbS_3$ that has previously been synthesized in the bulk (*16*, *17*). Atomic-resolution energy dispersive spectroscopy (EDS) shows that the 1D structures are composed of S and Nb with atomic ratio S/Nb = 3.23 ± 0.20, matching closely with the stoichiometry expected for $NbS_3$ (Fig. 1C and D). We note that there is ~1-5at% of Na and Cl in our samples as impurities. This EDS characterization also reveals that, post-growth, the particle contains Au, Nb, S and, notably, Na and Cl (Fig. SI3). This observation suggests a second role of NaCl due to its presence in the VLS catalyst, in addition to its use as an evaporation promoter as reported in the literature (*41*, *42*). We discuss below the role of NaCl in forming an alloyed VLS catalyst that guides the growth of 1D structures.

Figures 1E and F highlight the two morphologies of 1D materials that we typically see in our synthesis: one type is straight with an approximately square cross-section (referred to as nanowires (NWs), Fig. 1E) while the other has a ribbon morphology and rectangular cross section (referred to as nanoribbons (NRs), Figs. 1F-H). We note that Figs. 1E and F are representative of our synthesized materials regardless of the substrate, either $SiO_x$/Si or few-layer graphene suspended on a SiN TEM grid (Fig. S1C). The growth is reproducible: in a typical 30 minute-growth using 10 nm Au, more than 90% of the Au droplets have nucleated a nanostructure. The cross-sectional area of $NbS_3$ nanowires does not change greatly along the nanowire. In contrast, $NbS_3$ nanoribbons are more variable. Their cross section is rectangular (Figs. 1F-H), with non-planar sidewalls and growth direction varying along the length. A typical nanoribbon has a uniform width throughout its length (Fig. 1F), while we also observe other nanoribbons that show tapering (Figs. 1G and H) or a non-uniform cross-section (Fig. 1H). Moreover, many nanoribbons are thin and do not remain attached to the substrate. Some curl up towards the free-end tip, resembling a spiral structure (Fig. 1F). Similar "topological crystals" (*43*, *44*) have been observed in bulk $NbSe_3$ and $TaSe_3$ synthesized by chemical vapor transport. Interestingly, such structures are theoretically predicted to be energetically stable under certain nonequilibrium conditions, such as in the presence of Se droplets during growth (*43*). Measurements of the cross-section areas, lengths, and diameters of different $NbS_3$ morphologies can be found in Fig. S4.

Atomic-resolution imaging provides more insight into nanowire and nanoribbon morphology. In Fig. 2, high-angle annular dark-field (HAADF) STEM images show two distinct growth orientations, [100] (interchain) in Fig. 2A and [010] (intrachain) in Fig. 2B, in which the chain direction is parallel or perpendicular to the material growth direction, respectively (see atomic models in Fig. S1B). We will therefore refer to the nanostructures as b-nanowires (b-NW) and a-nanoribbons (a-NR) respectively, following the crystallographic convention that b is the chain direction in the crystal structure. Viewing the structure along the b direction, as in Fig. 2C, visualizes the bonding of neighbor chains into bilayers, a signature of TMT materials (*3*, *15*).

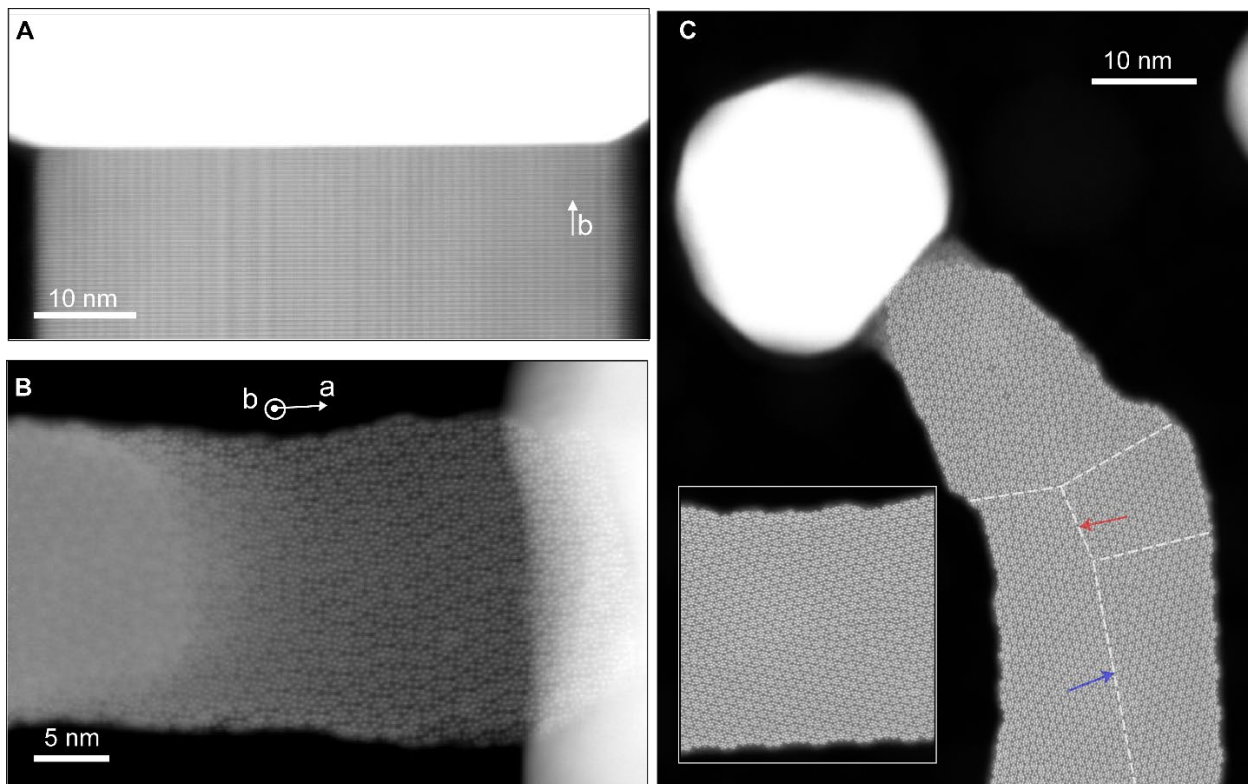

**Fig. 2. Atomic structure of VLS NbS₃ 1D structures.** (A-B) Atomic-resolution HAADF STEM images of a typical (A) b-nanowire (b-NW) and (B) a-nanoribbon (a-NR), showing the catalyst (bright region). The growth orientations are indicated by the white arrows. (C) Bilayer stacking disorder in a-NR. Rotation and rigid body shift between adjacent bilayer domains, indicated by the red and blue arrows, respectively, result in boundaries, some of which are displayed as the white dashed lines within the structure. (Inset): Higher magnification image of NbS₃ chains showing ordered regions and stacking disorder. Further examples of stacking disorder are shown in Fig. S5.

A prominent feature of both types of nanostructures is the variation in the stacking sequence and orientation of the bilayers. The stacking defects are clearly displayed in a-NRs oriented so that the prismatic chains point in the beam direction (Figs. 2B and C and Fig. S5). This end-on view provides a unique perspective of how the bilayers stack and arrange. In few-nm domains the bilayers stack in the same orientation with a rotation and/or rigid body shift between neighboring domains. In b-NWs, the disorder is visible in the variable spacing between the vertical lines (dark in TEM imaging, bright in STEM) in Figs. 1B and 2A, and the streaking in electron diffraction shown in the inset of Fig. 1B. The different crystal phases of NbS₃ are known to form depending on the growth conditions (*16*, *18*). The structure of the self-assembled NRs suggests that this may be in part attributed to the fact that the bilayer arrangement appears prone to stacking defects such as rotations and rigid body shifts, especially if new material is being added along the side of a series of chains. Changes in stacking order introduced during growth may eventually result in the different polymorphs reported in the literature (*16*, *18*).

The outline for the rest of our report is as follows. We first present strategies to control the dimensions (length and diameter) and morphologies of salt-assisted VLS-grown NbS₃. We then demonstrate the growth of NbSe$_3$ and TiS$_3$ using a similar approach. Finally, using structural correlations, we propose growth mechanisms for the different morphologies observed in the TMT nanostructures.

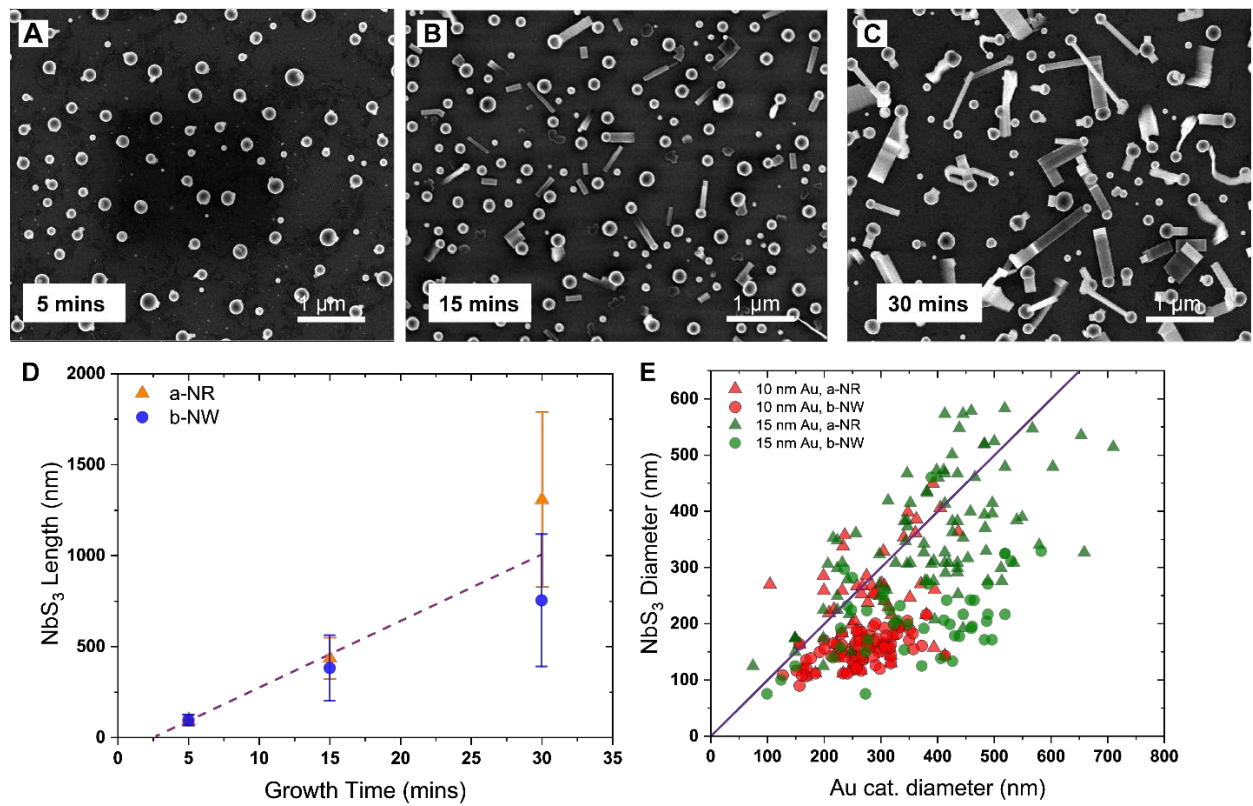

**Fig. 3. Size control of 1D NbS$_3$ materials.** (A-C) SEM images of NbS$_3$ growth using 10-nm Au catalyst film for growth times of 5, 15 and 30 minutes. (D) a-NR and b-NW length as a function of growth time. For each growth, 50-150 nanostructures were measured, exclusive of droplets without visible growth. The dash purple line indicates a linear fit to extract the growth rate. (E) Diameter of a-NR and b-NW as a function of Au diameter for 10-nm and 15-nm-thick Au (the thicknesses that produced the highest nanostructure yield). For a-NRs with a rectangular cross section, the longer side is used, as defined in the Fig. S4A schematics of a-NR and b-NW that indicate length, diameter and thickness. The purple line represents a 1:1 ratio of catalyst/nanowire diameters.

**Size control of 1D NbS$_3$ structures**

We find that the nanostructure length can be controlled through the growth time. Figures 3A-C show growths using 10-nm Au films on SiO$_x$/Si substrates for different periods of time with post-growth SEM showing the length of the nanostructures (Fig. 3D). The catalyst sizes for the three growth times are comparable (250 ± 120 nm), and are similar to droplet sizes after heating in the absence of the growth flux (Fig. S4A). This is consistent with dewetting kinetics of a uniform film where rapid initial dewetting is followed by a slower-changing configuration (*45*). After 5 minutes we see short nanowires (90 ± 30 nm) emerging from several catalyst particles. At this stage, it is hard to distinguish between a-NRs and b-NWs based on our SEM resolution. After 15 minutes, almost all catalysts have formed a nanostructure, the average length is 420 ± 140 nm, and one can start visually differentiating between a-NRs and b-NWs. For 30-min growth (a typical growth time for data presented in this study), a-NRs are longer with an average length of 1300 ± 480 nm, and b-NW are shorter at 650 ± 360 nm. A linear fit to length data (Fig. 3D) suggests an incubation time before growth begins of ~2.5 minutes. After that period, the nanostructures elongate linearly with a rate of 29.0 ± 8 nm/min for a-NRs and 25.0 ± 12 nm/min for b-NWs. We note that the deviation in the measurement might partially come from the tilt of the structures with respect to the imaging direction.

The diameter of the 1D $NbS_3$ materials can also be controlled with catalyst droplet size. Figure 3E shows the diameter of individual a-NRs and b-NWs as a function of their catalyst diameter for two growths that used 10-nm and 15-nm Au films. These films dewet to produce droplets with different but overlapping size distributions (45). Other growth conditions such as precursor mass, growth temperature, and growth time are kept the same. It appears that for both growths, the a-NRs have diameters comparable to their droplet diameter. This relationship is typical amongst VLS nanowires grown from other materials (25). On the other hand, the diameter of b-NWs is smaller than their catalyst diameter; the ratio of catalyst/nanowire diameters for b-NW in both cases is ~1.8. It is worth noting that the a-NRs observed here are distinctive compared to what has been reported in nano-ribbon structures of oxides (46), such as ZnO and $SnO_2$, and layered materials (47), for example $Bi_2Se_3$ and GaSe. In these materials, the catalyst diameter is often much smaller than the width of the ribbons and the droplets reside at one corner of the ribbon's top surface (somewhat like the structure in Fig. 1H). The growth mechanisms of these ribbon structures are therefore often attributed to a mixture of both VLS and vapor-solid (VS) mechanisms (46, 47) with VS dominating and creating highly tapered structures. We will compare the growth mechanism of our $NbS_3$ a-NRs with these reports in the subsequent discussion.

**Morphology control of 1D NbS₃ structures: nanowires vs. nanoribbons**

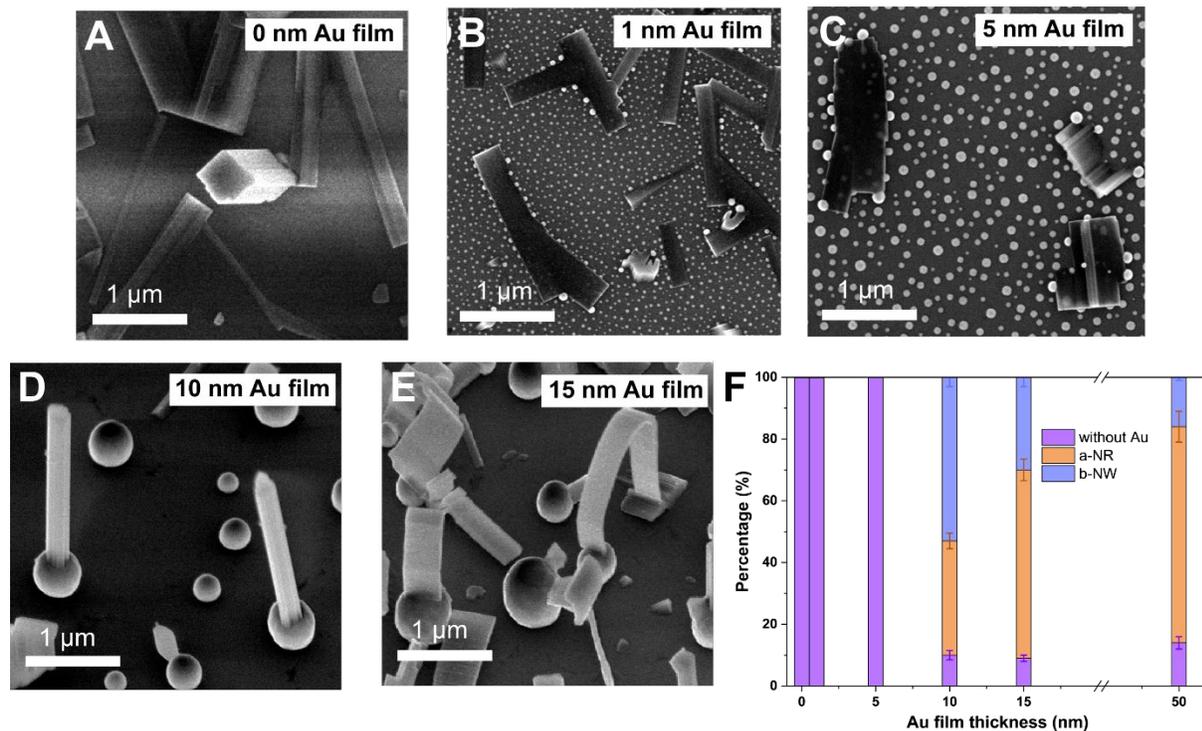

**Fig. 4. Relationship between Au thickness and NbS₃ morphology.** (A-E) SEM images of NbS₃ structures synthesized using Au thin films with different thickness (0, 1, 5, 10 and 15 nm) for the same growth time of 30 minutes and growth temperature of 725°C. For 50-nm Au, refer to the patterned growth in Fig. S1B. (F) Population percentage of 1D structures without Au catalyst particle, a-NR and b-NW observed in the growths using different Au film thicknesses.

We now investigate the role of Au in determining the growth morphology, particularly whether VLS growth takes place at all and whether the structure is an a-NR or b-NW. We find that the deposited Au thickness, or equivalently, the catalyst size (*45*), does affect the nanostructure type (Fig. 4). Figure 4A shows growth on SiO$_x$/Si substrates without Au catalyst (0 nm Au film) results in mostly b-NWs. This non-catalyzed growth mode is shown in more detail in Fig. S6. 1-nm and 5-nm Au films similarly resulted in *no* VLS nanowires (Figs. 4B and C). NbS₃ nanostructures are visible but are broad and flat, similar to what we observe in growth without Au (Fig. 4A), and are decorated by several small droplets around their edges instead of having a single particle at one end as in the typical VLS nanowires in Figs. 1 and 2. This suggests that for thin Au (1 and 5 nm), the NbS₃ 1D structures nucleate spontaneously and Au particles attach to them. With thicker Au (10, 15, 50 nm), the grown materials show the telltale signatures of VLS growth described above. Furthermore, the nanostructure type is correlated with the Au film thickness. The fraction of the population that consists of a-NR, b-NW and uncatalyzed growth in experiments using different Au film thicknesses is shown in Fig. 4F. As the film thickness increases, a-NRs become dominant. This trend is most striking at 50-nm Au, where a-NRs make up 70% of the grown nanowires. In Fig. 4E, we show that on a single substrate a-NRs and b-NWs grow from an overlapping distribution of Au droplets, but there is a trend for the larger droplets to produce more a-NRs. The increasing proportion of a-NRs in thicker Au films is consistent with this observation, since thicker films dewet into droplets that are larger on average. Overall, we suggest that the possibility exists for controlling the 1D morphologies of $NbS_3$ by choice of the Au film thickness or by using catalysts of well-defined initial size.

**Materials control: Salt-assisted VLS synthesis of other 1D transition metal trichalcogenides**

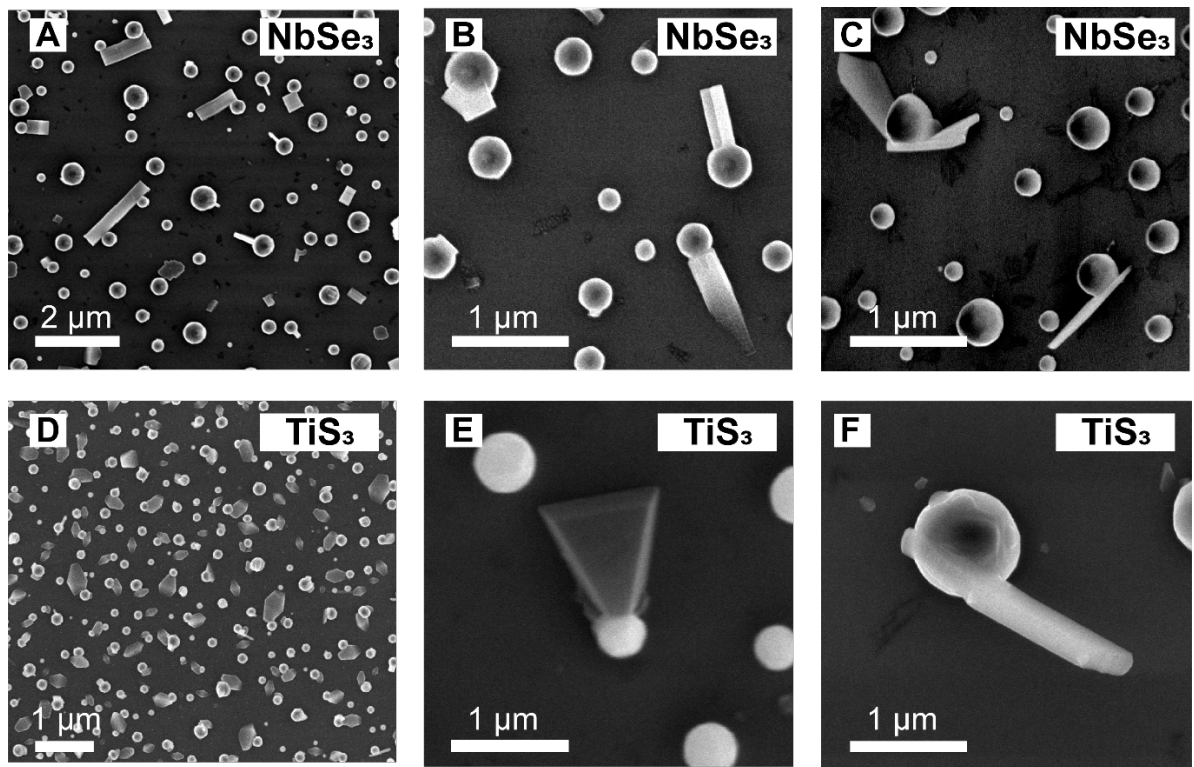

**Fig. 5. Salt-assisted VLS growth of various TMTs**. SEM images showing 1D nanomaterials observed in the salt-assisted VLS growth of (A-C) $NbSe_3$ and (D-F) $TiS_3$.

We extend our salt-assisted VLS method to synthesize nanostructures in two other vdW chain materials, $NbSe_3$ and $TiS_3$, to explore the generality of the approach to a variety of transition metals (Nb, Ti) and chalcogens (S, Se). Figure 5 shows representative SEM images of $NbSe_3$ and $TiS_3$ 1D structures with SEM-EDS elemental mapping shown in Fig. S7. We note that for $NbSe_3$, the yield of 1D materials at the optimum $NbS_3$ growth temperature of 700-750°C is lower than in the growth of $NbS_3$. To increase the density of $NbSe_3$ nanostructures the substrate temperature was raised to the 800-850°C range. The result is $NbSe_3$ with nanowire morphology (Fig. 5b) with [010] growth orientations, similar to $NbS_3$ and consistent with the similarity of the $NbSe_3$ and $NbS_3$ crystal structures. However, we also observe plate-like half-hexagon structures stemming from a single Au catalyst (Fig. 5C). We attribute this 2D-like structure to the presence of the layered hexagonal material $NbSe_2$, since at the high growth temperature, $NbSe_2$ is known (*48*, *49*) to be more stable than $NbSe_3$. This is analogous to the presence of $NbS_2$ at high growth temperatures discussed above and in Fig. S2. $TiS_3$ synthesized under the same condition as $NbS_3$ (Figs. 5D-F) displays a mix of 1D wire and truncated trapezoid morphology. By analogy we suggest that at 700-750°C there exists a mix of 1D $TiS_3$ and 2D $TiS_2$ (*8*, *50*), which result in structures having mixed features of wire-like (1D) and plate-like (2D) morphologies. Optimization of the synthesis parameters, mostly the interplay between the temperature of the substrate and precursors and even the choice of alkaline metal halides (*51*) should offer prospects for improving the density of $NbSe_3$ nanowires and increasing the proportion of the 1D phase of $TiS_3$.

**The mechanisms at work during salt-assisted VLS growth**

To understand potential growth mechanisms that could describe the above results, we consider the relationships between the source materials, growth parameters and morphologies of $NbS_3$ 1D nanostructures. We first discuss the possible role of NaCl in VLS growth. We note that our growth is the first to use *both* NaCl and Au, for any metal chalcogenide (including mono-, di- and tri-chalcogenide) growth by CVD. NaCl alone has been shown as an effective growth promoter in the synthesis of 2D vdW materials such as $MoS_2$, $WSe_2$ and $NbSe_2$ (*41*, *42*). In these cases, it was suggested that NaCl reacts with transition metal precursors, for instance the metal and its oxide (e.g., $MoO_3$, W, $Nb_2O_5$), to create volatile intermediate compounds (metal oxychlorides) with much lower melting points, and thus, higher vapor pressures at a given growth temperature compared to the metal or metal oxide precursors alone (*35*, *36*). For example, the possible reaction route for the case of $WSe_2$ is (*42*):

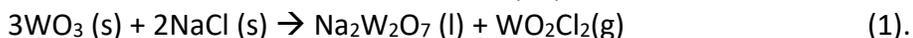

$$3WO_3 \text{ (s)} + 2NaCl \text{ (s)} \rightarrow Na_2W_2O_7 \text{ (l)} + WO_2Cl_2\text{(g)} \qquad (1).$$

The nucleation and growth of $WSe_2$ take place by a dissociation/formation reaction to replace W-O and W-Cl bonds at the step edges of the material by W-Se bonds, as suggested recently in Ref. (*51*):

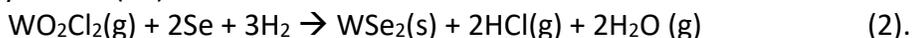

$$WO_2Cl_2\text{(g)} + 2Se + 3H_2 \rightarrow WSe_2\text{(s)} + 2HCl\text{(g)} + 2H_2O \text{ (g)} \qquad (2).$$

In our growth of $NbS_3$, we postulate that similar reactions between the NaCl and Nb (presumably with an oxide shell, $Nb_2O_5$, since the source was prepared and growth carried out at ambient pressure (*52*)) generate metal oxychloride products, such as $NbOCl_3$ which has a moderately low melting point of 200°C (*53*). These volatile species are then transported to the growth substrate simultaneously with the S vapor (see the growth setup in Fig. S1) to initiate the formation of the final structure. A control experiment using only Nb and S results in little or no $NbS_3$ formation, suggesting that virtually no evaporation of the Nb precursor takes place at 600-850°C. This supports the role of NaCl as an evaporation promoter.

However, our EDS results mentioned earlier and shown in Fig. S3 indicate a second potential role of NaCl in our VLS growth of $NbS_3$ nanowires. We find that post-growth the catalyst particle contains not only Au, Nb, S, but also Na and Cl. The Na and Cl concentrations are comparable to those of the growth species, suggesting that this is not merely a small contamination of NaCl arising from its role as a growth precursor. This observation may help explain why, to the best of our knowledge, there has not previously been a synthesis of any transition metal chalcogenide nanowire (including mono, di and tri-chalcogenides) by VLS. VLS growth generally requires the formation of a eutectic between the catalyst and the nanowire materials. However, Au does not form eutectic droplets with Nb or S (based on the individual binary phase diagrams) below 1064°C, the melting point of Au. The lack of a reactive Au compound to act as a VLS catalyst has been demonstrated in the literature for some growths of transition metal dichalcogenide materials, for example $MoS_2$ and $WS_2$, where Au only serves as a growth substrate (*54*) or nucleation point (*55*) rather than directly participating in the growth process. Moreover, Au foil and deposited Au triangles in the presence of transition metals (Mo, W) and chalcogen (S) vapors in these studies (at a similar temperature range of 650-750°C to that used in our synthesis) do not react, instead remaining in their original morphologies (*55*) instead of transforming into spherical particles, as we observed (post-growth) in our $NbS_3$ experiments. In this regard, we hypothesize that the catalyst droplet forms as a eutectic composition of Au and Na/Cl-containing intermediates at our growth temperature of ~725°C. We suggest that NaCl vapor (abundant at the growth temperature of 725-750°C) is mobile under the growth conditions and is able to be incorporated into the Au nanostructures. We note that the post-growth spherical shape of the catalyst particles and observations we have made of adjacent particles that have merged could imply a liquid eutectic state during the growth.

We therefore propose that NaCl in our growth plays two roles. It serves as an evaporation promoter that reacts with Nb precursors to create volatile metal oxychloride intermediates (one possibility being $NbClO_3$), which enhances the mass flux of precursors to enable or accelerate growth at low temperatures (650-750°C). We also suggest that NaCl vapor may adsorb on Au and form a eutectic alloy, creating a reactive liquid surface at which Nb vapor compound(s), such as $NbClO_3$, and S are incorporated after arriving from the vapor phase to result in $NbS_3$ nanowires via the VLS mechanism.

We now consider the morphologies produced as growth proceeds. In VLS, the growth species have two pathways (*56*) by which adatoms incorporate into the growth front of a nanowire. The first pathway, which is unique to VLS growth, is adatom absorption through the catalyst droplet, assumed due to a higher sticking coefficient on the droplet surface compared to the substrate; this drives supersaturation of the growth species in the droplet, and therefore subsequent nucleation and growth of atomic layers at the catalyst-nanowire interface (*22*). To validate that this VLS pathway is indeed active in the formation of our nanostructures, we performed an interrupted growth in which a TEM-compatible substrate (Fig. S8A) was imaged directly after growth and then returned to the furnace for a second growth under the same conditions (Fig. S8B). Comparison of the images for one $NbS_3$ b-NW (marked by the red rectangle) showed that growth took place at the catalyst-nanowire interface (Fig. S8C) as expected for the VLS mechanism, rather than, for example, at the non-Au tip of the nanowire.

The second pathway involves either direct impingement onto the sidewall or adatom diffusion from the substrate to some part of the nanowire, followed by diffusion along the nanowire sidewalls and incorporation at any part of the nanowire surface, resulting in growth via the vapor-solid (VS) mechanism (*25*). This type of surface diffusion pathway is similar to the growth mechanism suggested in salt-assisted synthesis of 2D transition metal dichalcogenide materials (*38*, *51*), in which $MoS_2$ grows quickly by reaction and attachment of adatoms preferentially at the step edges. Since the overall geometry and nature of the surface are important parameters in the VS process, we might expect the VS pathway to have a different overall effect on a-NRs and b-NWs. These nanostructures exhibit surfaces with different attachment sites and diffusion parameters, due to the relationship between the anisotropic crystal structure and the growth direction. For b-NW, with growth orientation [010] or intrachain orientation, the available surface for transporting adatoms is along the chains. For a-NRs, with growth orientation [100] or interchain orientation, the surfaces present are either parallel to the chains or composed of a perpendicular plane that cuts through chains. b-NWs generally show flat step-free surfaces, suggesting that diffusion along these planes is fast and attachment is difficult unless steps are present. a-NRs show rougher surfaces implying that attachment is more favorable. This may therefore imply that a-NRs could be growing with a proportionally larger contribution from VS compared to VLS growth. This is consistent with the greater variety in shape and frequent tapering visible in a-NRs, as shown in Figs. 1F-H. Indeed, a general observation, when considering the contribution of VS to a VLS process, is that each nanostructure may be affected differently, due to the variable geometry of the sidewalls present. Here, some a-NRs crawl along the surface while others grow off the surface (Fig. 1). This may account for the scatter in data for dimension and growth rates in Figs. 3D and S4, even for similar catalyst diameters and hence VLS growth rates.

We finally discuss the factors that determine whether growth will produce a-NR or b-NW morphologies. In VLS growth (in the most general case, say where droplets are present on an amorphous substrate), the initial nucleation event is the formation of a small crystallite either at the triple phase line or at the droplet surface (*57, 58*). The shape of this nucleus depends on the energies of the interfaces between the growth material and droplet material, growth material and substrate, and growth material and vapor phase. These parameters are generally unknown, but in growths without Au catalysts (Fig. 4A), we observe structures that appear more like b-NWs (Fig. S6). This suggests that in $NbS_3$ the planes that are parallel to the chain directions may have lower energies. If the initial nucleus follows these energetic considerations it is likely to be elongated in the b direction. Needle-shaped nuclei lying at the droplet surface will create a-NR structures as material is added to the planes parallel to the chain directions. However, if the nucleus rotates (*57*) at all, then elongation along the b direction becomes impossible and a b-NW is more likely to result. Geometrically we might therefore conclude that a larger droplet with lower surface curvature favors the orientation of needle shaped nuclei parallel to the droplet surface and hence the growth of a-NRs. This is consistent with our finding that thicker Au films produce more a-NRs compared to the number of b-NW. This simple argument neglects the complexities of the energetics and kinetics; further experiments with well-defined catalyst geometry would help to explore these possibilities.

Based on these discussions we propose that the salt-assisted VLS synthesis of $NbS_3$ 1D nanostructures operates simultaneously with a VS mechanism that appears to contribute relatively more to the growth of a-NRs. We suggest that the morphology that is produced depends on the geometry present at the nucleation event, and this may depend on the anisotropy in surface energies and growth rates for the nucleus. Further studies with size-selected Au nanoparticles, rather than the dewetted films used here, will enable exploration of the specifics of how the catalyst-related VLS and attachment/diffusion-related VS pathways favor the formation of a-NR versus b-NW morphologies.

## Conclusions

We have presented a growth strategy that uses an alloy from Au and NaCl as a catalyst to synthesize a variety of 1D vdW materials, $NbS_3$, $NbSe_3$ and $TiS_3$, on bulk substrates ($SiO_x$/Si, SiN, sapphire, mica) as well as on van der Waals surfaces (graphene, hexagonal BN). The method we describe, salt-assisted VLS growth, combines key benefits of both vapor-liquid-solid and salt-assisted growth processes. We show that the thickness of the Au film that dewets to form the catalytic droplets and the positions at which it is patterned onto the substrate offer the ability to control the size and location of the structures, as is the case in a conventional VLS process. The presence of salt offers advantages of fast growth at moderate temperatures (650-750°C), producing a high yield of $NbS_3$ nanostructures with sub-ten nanometer diameter and lengths up to several tens of micrometers within a relatively short time (15-45 minutes) compared to growth in the absence of salt. For $NbS_3$ we describe two distinct morphologies that result from this growth mechanism: b-NWs, straight nanowires in which the growth direction s along the chain direction ([010]), and a-NRs, where dimensions are more variable and growth is perpendicular to the chain direction ([100]). We outline the key synthesis parameters, particularly the growth time and the thickness of the initial Au film, that influence the dimensions and morphologies of these different $NbS_3$ 1D structures. We show in particular that thicker Au (50 nm) results in more a-NRs, while thinner initial Au (10 nm) favors the growth of b-NWs. The diameter of b-NWs can be constrained within a small range (80-100 nm) by controlling the Au layer thickness and hence the droplet size after Au dewetting. We demonstrate the generality of salt-assisted VLS growth by creating nanowire morphologies in other 1D transition metal trichalcogenides, $NbSe_3$ and $TiS_3$. We explored the details of the salt-assisted growth mechanism through the relationship between morphology and growth parameters. We propose the coexistence of two pathways, catalyst limited growth and surface diffusion limited growth (VLS and VS), to account for the difference in the growth rates, geometry and relative frequency of $NbS_3$ a-NRs and b-NWs.

We believe that our demonstration of this new growth mechanism for nanowires and nanoribbons of transition metal trichalcogenides ($NbS_3$, $NbSe_3$ and $TiS_3$) will expand the ability to grow emerging 1D vdW materials, including their heterostructures, and make use of their properties. Additionally, we propose that our method will also create opportunities to explore nanowire synthesis and nanowire heterostructures in group IV and V heavy-metal compounds (e.g., oxides of Mo, W, Ta, Hf). The synthesis of oxide nanowires of Mo, W, Ta, Hf based on vapor-phase synthesis, including VLS, is challenging, mainly because of the difficulty of creating and supplying sufficient vapor pressure of the metal precursors. The results obtained for $NbS_3$ shows this challenge can be overcome with the presence of salt under appropriate conditions; in this way the reactants can be incorporated into a catalyst particle, as in the case of conventional VLS, to direct the growth of nanowires. Further measurements are required to explore the range of applicability of the salt-assisted VLS process.

Going forward, it is important to measure the properties of the grown nanostructures to establish any effects of the catalytic elements present. It will then be possible to understand how the native anisotropy of these synthesized TMT materials governs properties such as conductivity and linear dichroism, and how the defects we see within the self-assembled nanostructures affect their electrical and thermal transport properties. The capability we have demonstrated, to create TMT nanostructures with some control of diameter and growth location, opens opportunities for such studies. We anticipate a greater understanding of the basic science of this versatile class of materials and technological advances as components of quantum devices where precise placement of individual identical nanowires is required (*59*), for example in nano-thermoelectrics or in fault-tolerant quantum circuits.

**MATERIALS AND METHODS**

**Substrate preparation:** $SiO_x$/Si, SiN, a-sapphire and mica bulk substrates are cleaned with isopropanol and deionized water, followed by oxygen plasma (150 mTorr, 5 minutes). To create 2D substrates, few-layer graphene and hexagonal boron nitride were first mechanically exfoliated by a Scotch-tape method, then transferred onto holey SiN TEM grids. Au films are then e-beam evaporated onto the substrates with different thickness (1, 3, 5, 10, 15, 30, 50 nm).

**CVD growth:** 50 mg of Nb powder is mixed with 25 mg of NaCl (Sigma Aldrich), and the mixture is loaded into an alumina boat. A growth substrate is placed upside down right on top of the precursor mixture and the Nb/NaCl/substrate boat is set to a temperature of 650-800 °C. 150-200 mg of S (Sigma Aldrich) is put upstream and far away from the hot zone. S boat is kept at room temperature in the beginning of each growth and it is only pushed into a hot zone (130-150 °C) only after the Nb/NaCl/substrate boat reaches the target temperature. 100-150 sccm of high purity Ar (Airgas) was used as the carrier gas. The growth time varies from 5 to 45 minutes. At the end of the growth, the carrier gas was turned up to 200 sccm and the tube furnace's cap was open to cool down the furnace quickly. For growths on 2D materials on SiN TEM grid, we machine a boron nitride holder which allows to load a TEM grid upside down to maintain the same growth setup as in the growths using a bulk substrate. For the growths of $NbSe_3$ and $TiS_3$, we used the same growth parameters, except $T_{growth} = 800°C$ for $NbSe_3$ and 50 mg of $TiO_2$ for $TiS_3$ synthesis.

**Material characterization:** SEM imaging was carried out at 5-20 kV on FEI Helios Nanolab 600. S/TEM characterization was carried out in a Themo Fisher Scientific Themis Z STEM operated at 200 kV. Typical aberration-corrected STEM imaging conditions are 19 mrad semi-convergent angle (estimated sub-angstrom probe size) and 80 pA probe current. EDS was collected at higher probe current of 120-200 pA. Elemental mapping and quantification used Thermo Fisher Velox software.

### Acknowledgements

We acknowledge funding from the U.S. Department of Energy, Office of Basic Energy Sciences, Division of Materials Sciences and Engineering under Award DE-SC0019336. This work made use of the Shared Experimental Facilities supported in part by the MRSEC program of the National Science Foundation under award number DMR-1419807. The authors thank Dr. Aubrey Penn at MIT.nano for assistance with STEM characterization.

**Author contributions:** Conceptualization: T.P. and F.M.R. Methodology: T.P and F.M.R. Formal analysis: T.P., M.A.F and F.M.R. Investigation, experiments, validation: T.P, K.R, J.D.H, B.W., N.D. Data curation: T.P. and F.M.R. Writing-original draft: T.P., M.A.F and F.M.R. Writing-review and editing: all authors. Visualization: T.P. Supervision: M.A.F. and F.M.R. Funding acquisition: M.A.F. and F.M.R.

**Competing interests:** The authors declare that they have no competing interests.

**Data and materials availability:** All data needed to evaluate the conclusions in the paper are present in the paper and/or the Supplementary Materials.

### Supplementary Materials

This PDF file includes:
Supplementary Text

Figs. S1 to S8

## References

1. K. S. Novoselov, Two-dimensional atomic crystals. *Proc. Natl Acad. Sci. USA*. **102** (2005).

2. A. K. Geim, I. V. Grigorieva, Van der Waals heterostructures. *Nature*. **499**, 419–425 (2013).

3. A. A. Balandin, F. Kargar, T. T. Salguero, R. K. Lake, One-dimensional van der Waals quantum materials. *Materials Today*. **55**, 74–91 (2022).

4. R. V. Coleman, B. Giambattista, P. K. Hansma, A. Johnson, W. W. McNairy, C. G. Slough, Scanning tunnelling microscopy of charge-density waves in transition metal chalcogenides. *Adv Phys*. **37**, 559–644 (1988).

5. A. Patra, C. S. Rout, Anisotropic quasi-one-dimensional layered transition-metal trichalcogenides: synthesis, properties and applications. *RSC Adv*. **10**, 36413–36438 (2020).

6. J. O. Island, A. J. Molina-Mendoza, M. Barawi, R. Biele, E. Flores, J. M. Clamagirand, J. R. Ares, C. Sánchez, H. S. J. Van Der Zant, R. D'Agosta, I. J. Ferrer, A. Castellanos-Gomez, Electronics and optoelectronics of quasi-1D layered transition metal trichalcogenides. *2d Mater*. **4** (2017), p. 000000.

7. T. Pham, S. Oh, P. Stetz, S. Onishi, C. Kisielowski, M. L. Cohen, A. Zettl, Torsional instability in the single-chain limit of a transition metal trichalcogenide. *Science (1979)*. **361**, 263–266 (2018).

8. J. O. Island, M. Barawi, R. Biele, A. Almazán, J. M. Clamagirand, J. R. Ares, C. Sánchez, H. S. J. van der Zant, J. V. Álvarez, R. D'Agosta, I. J. Ferrer, A. Castellanos-Gomez, TiS3 Transistors with Tailored Morphology and Electrical Properties. *Advanced Materials*. **27**, 2595–2601 (2015).

9. V. E. Fedorov, S. B. Artemkina, E. D. Grayfer, N. G. Naumov, Y. V. Mironov, A. I. Bulavchenko, V. I. Zaikovskii, I. V. Antonova, A. I. Komonov, M. V. Medvedev, Colloidal solutions of niobium trisulfide and niobium triselenide. *J Mater Chem C Mater*. **2**, 5479–5486 (2014).

10. T. A. Empante, A. Martinez, M. Wurch, Y. Zhu, A. K. Geremew, K. Yamaguchi, M. Isarraraz, S. Rumyantsev, E. J. Reed, A. A. Balandin, L. Bartels, Low Resistivity and High Breakdown Current Density of 10 nm Diameter van der Waals TaSe3 Nanowires by Chemical Vapor Deposition. *Nano Lett*. **19**, 4355–4361 (2019).

11. L. Yang, Y. Tao, Y. Zhu, M. Akter, K. Wang, Z. Pan, Y. Zhao, Q. Zhang, Y.-Q. Xu, R. Chen, T. T. Xu, Y. Chen, Z. Mao, D. Li, Observation of superdiffusive phonon transport in aligned atomic chains. *Nat Nanotechnol*, 1–5 (2021).

12. S. Meyer, T. Pham, S. Oh, P. Ercius, C. Kisielowski, M. L. Cohen, A. Zettl, Metal-insulator transition in quasi-one-dimensional HfTe3 in the few-chain limit. *Phys Rev B*. **100**, 041403 (2019).

13. T. Pham, S. Oh, S. Stonemeyer, B. Shevitski, J. D. Cain, C. Song, P. Ercius, M. L. Cohen, A. Zettl, Emergence of Topologically Nontrivial Spin-Polarized States in a Segmented Linear Chain. *Phys Rev Lett*. **124**, 206403 (2020).

14. S. G. Zybtsev, V. Y. Pokrovskii, V. F. Nasretdinova, S. V. Zaitsev-Zotov, V. V. Pavlovskiy, A. B. Odobesco, W. W. Pai, M. W. Chu, Y. G. Lin, E. Zupanič, H. J. P. Van Midden, S. Šturm, E. Tchernychova, A. Prodan, J. C. Bennett, I. R. Mukhamedshin, O. V. Chernysheva, A. P.

Menushenkov, V. B. Loginov, B. A. Loginov, A. N. Titov, M. Abdel-Hafiez, Nb S3: A unique quasi-one-dimensional conductor with three charge density wave transitions. *Phys Rev B*. **95**, 035110 (2017).

15. S. Conejeros, B. Guster, P. Alemany, J. P. Pouget, E. Canadell, Rich Polymorphism of Layered NbS3. *Chemistry of Materials*. **33**, 5449–5463 (2021).

16. M. A. Bloodgood, P. Wei, E. Aytan, K. N. Bozhilov, A. A. Balandin, T. T. Salguero, Monoclinic structures of niobium trisulfide. *APL Mater*. **6**, 026602 (2017).

17. E. Zupanič, H. J. P. van Midden, M. A. van Midden, S. Šturm, E. Tchernychova, V. Ya. Pokrovskii, S. G. Zybtsev, V. F. Nasretdinova, S. V. Zaitsev-Zotov, W. T. Chen, W. W. Pai, J. C. Bennett, A. Prodan, Basic and charge density wave modulated structures of NbS3-II. *Phys Rev B*. **98**, 174113 (2018).

18. M. A. Bloodgood, Y. Ghafouri, P. Wei, T. T. Salguero, Impact of the chemical vapor transport agent on polymorphism in the quasi-1D NbS3 system. *Appl Phys Lett*. **120**, 173103 (2022).

19. M. A. Stolyarov, G. Liu, M. A. Bloodgood, E. Aytan, C. Jiang, R. Samnakay, T. T. Salguero, D. L. Nika, S. L. Rumyantsev, M. S. Shur, K. N. Bozhilov, A. A. Balandin, Breakdown current density in h -BN-capped quasi-1D TaSe 3 metallic nanowires: prospects of interconnect applications. *Nanoscale*. **8**, 15774–15782 (2016).

20. R. S. Wagner, W. C. Ellis, Vapor-liquid-solid mechanism of single crystal growth. *Appl Phys Lett*. **4**, 89–90 (1964).

21. D. Jacobsson, F. Panciera, J. Tersoff, M. C. Reuter, S. Lehmann, S. Hofmann, K. A. Dick, F. M. Ross, Interface dynamics and crystal phase switching in GaAs nanowires. *Nature*. **531**, 317–322 (2016).

22. F. M. Ross, Controlling nanowire structures through real time growth studies. *Reports on Progress in Physics*. **73**, 114501 (2010).

23. Z. L. Wang, Zinc oxide nanostructures: Growth, properties and applications. *Journal of Physics Condensed Matter*. **16**, R829 (2004).

24. T. Pham, S. Kommandur, H. Lee, D. Zakharov, M. A. Filler, F. M. Ross, One-dimensional twisted and tubular structures of zinc oxide by semiconductor-catalyzed vapor–liquid–solid synthesis. *Nanotechnology*. **32**, 075603 (2020).

25. M. Ek, M. A. Filler, Atomic-Scale Choreography of Vapor-Liquid-Solid Nanowire Growth. *Acc Chem Res*. **51**, 118–126 (2018).

26. E. Barrigón, M. Heurlin, Z. Bi, B. Monemar, L. Samuelson, Synthesis and Applications of III-V Nanowires. *Chem Rev*. **119** (2019), pp. 9170–9220.

27. L. Güniat, P. Caroff, A. Fontcuberta I Morral, Vapor Phase Growth of Semiconductor Nanowires: Key Developments and Open Questions. *Chem Rev*. **119** (2019), pp. 8958–8971.

28. E. Sutter, J. S. French, S. Sutter, J. C. Idrobo, P. Sutter, Vapor-Liquid-Solid Growth and Optoelectronics of Gallium Sulfide van der Waals Nanowires. *ACS Nano*. **14**, 6117–6126 (2020).

29. P. Sutter, J. S. French, L. Khosravi Khorashad, C. Argyropoulos, E. Sutter, Optoelectronics and Nanophotonics of Vapor-Liquid-Solid Grown GaSe van der Waals Nanoribbons. *Nano Lett*. **21**, 4335–4342 (2021).

30.  H. Peng, D. T. Schoen, S. Meister, X. F. Zhang, Y. Cui, Synthesis and phase transformation of In2Se3 and CuInSe2 nanowires. *J Am Chem Soc*. **129**, 34–35 (2007).

31.  J. J. Cha, J. R. Williams, D. Kong, S. Meister, H. Peng, A. J. Bestwick, P. Gallagher, D. Goldhaber-Gordon, Y. Cui, Magnetic doping and kondo effect in Bi2Se3 nanoribbons. *Nano Lett*. **10**, 1076–1081 (2010).

32.  P. Sutter, S. Wimer, E. Sutter, Chiral twisted van der Waals nanowires. *Nature*. **570** (2019), pp. 354–357.

33.  Y. Liu, J. Wang, S. Kim, H. Sun, F. Yang, Z. Fang, N. Tamura, R. Zhang, X. Song, J. Wen, B. Z. Xu, M. Wang, S. Lin, Q. Yu, K. B. Tom, Y. Deng, J. Turner, E. Chan, D. Jin, R. O. Ritchie, A. M. Minor, D. C. Chrzan, M. C. Scott, J. Yao, Helical van der Waals crystals with discretized Eshelby twist. *Nature*. **570** (2019), pp. 358–362.

34.  Z. Fang, Y. Liu, S. Gee, S. Lin, S. Koyama, C. So, F. Luo, R. Chen, B. Tang, J. Yao, Chemically Modulating the Twist Rate of Helical van der Waals Crystals. *Chemistry of Materials*. **32**, 299–307 (2020).

35.  J. Zhou, J. Lin, X. Huang, Y. Zhou, Y. Chen, J. Xia, H. Wang, Y. Xie, H. Yu, J. Lei, D. Wu, F. Liu, Q. Fu, Q. Zeng, C. H. Hsu, C. Yang, L. Lu, T. Yu, Z. Shen, H. Lin, B. I. Yakobson, Q. Liu, K. Suenaga, G. Liu, Z. Liu, A library of atomically thin metal chalcogenides. *Nature*. **556**, 355–359 (2018).

36.  S. Li, Y. C. Lin, W. Zhao, J. Wu, Z. Wang, Z. Hu, Y. Shen, D. M. Tang, J. Wang, Q. Zhang, H. Zhu, L. Chu, W. Zhao, C. Liu, Z. Sun, T. Taniguchi, M. Osada, W. Chen, Q. H. Xu, A. T. S. Wee, K. Suenaga, F. Ding, G. Eda, Vapour-liquid-solid growth of monolayer MoS2 nanoribbons. *Nat Mater*. **17**, 535–542 (2018).

37.  M. Chen, L. Li, M. Xu, W. Li, L. Zheng, X. Wang, Quasi-One-Dimensional van der Waals Transition Metal Trichalcogenides. *Research* (2023), doi:10.34133/RESEARCH.0066.

38.  J. Zhou, J. Lin, X. Huang, Y. Zhou, Y. Chen, J. Xia, H. Wang, Y. Xie, H. Yu, J. Lei, D. Wu, F. Liu, Q. Fu, Q. Zeng, C. H. Hsu, C. Yang, L. Lu, T. Yu, Z. Shen, H. Lin, B. I. Yakobson, Q. Liu, K. Suenaga, G. Liu, Z. Liu, A library of atomically thin metal chalcogenides. *Nature*. **556**, 355–359 (2018).

39.  H. Wang, X. Huang, J. Lin, J. Cui, Y. Chen, C. Zhu, F. Liu, Q. Zeng, J. Zhou, P. Yu, X. Wang, H. He, S. H. Tsang, W. Gao, K. Suenaga, F. Ma, C. Yang, L. Lu, T. Yu, E. H. T. Teo, G. Liu, Z. Liu, High-quality monolayer superconductor NbSe2 grown by chemical vapour deposition. *Nature Communications 2017 8:1*. **8**, 1–8 (2017).

40.  E. V. Formo, J. A. Hachtel, Y. Ghafouri, M. A. Bloodgood, T. T. Salguero, Thermal Stability of Quasi-1D NbS3Nanoribbons and Their Transformation to 2D NbS2: Insights from in Situ Electron Microscopy and Spectroscopy. *Chemistry of Materials*. **34**, 279–287 (2022).

41.  C. Xie, P. Yang, Y. Huan, F. Cui, Y. Zhang, Roles of salts in the chemical vapor deposition synthesis of two-dimensional transition metal chalcogenides. *Dalton Transactions*. **49**, 10319–10327 (2020).

42.  W. Han, K. Liu, S. Yang, F. Wang, J. Su, B. Jin, H. Li, T. Zhai, Salt-assisted chemical vapor deposition of two-dimensional materials. *Sci China Chem*. **62** (2019), pp. 1300–1311.

43.  A. N. Enyashin, A. L. Ivanovskiĭ, Electronic, energy, and thermal properties of the Möbius strip and related ring nanostructures of NbS3. *Physics of the Solid State 2006 48:4*. **48**, 780–785 (2006).

44.    M. Tsubota, K. Inagaki, T. Matsuura, S. Tanda, Polyhedral topological crystals. *Cryst Growth Des*. **11**, 4789–4793 (2011).

45.    C. V. Thompson, Solid-State Dewetting of Thin Films. *Annual Reviews*. **42**, 399–434 (2012).

46.    Z. L. Wang, Splendid one-dimensional nanostructures of zinc oxide: A new nanomaterial family for nanotechnology. *ACS Nano*. **2**, 1987–1992 (2008).

47.    D. Kong, J. C. Randel, H. Peng, J. J. Cha, S. Meister, K. Lai, Y. Chen, Z. X. Shen, H. C. Manoharan, Y. Cui, Topological insulator nanowires and nanoribbons. *Nano Lett*. **10**, 329–333 (2010).

48.    Y. S. Hor, U. Welp, Y. Ito, Z. L. Xiao, U. Patel, J. F. Mitchell, W. K. Kwok, G. W. Crabtree, Superconducting NbSe2 nanowires and nanoribbons converted from NbSe3 nanostructures. *Appl Phys Lett*. **87**, 142506 (2005).

49.    M. Nath, S. Kar, A. K. Raychaudhuri, C. N. R. Rao, Superconducting NbSe2 nanostructures. *Chem Phys Lett*. **368**, 690–695 (2003).

50.    C. G. Hawkins, L. Whittaker-Brooks, Controlling Sulfur Vacancies in TiS2-x Cathode Insertion Hosts via the Conversion of TiS3 Nanobelts for Energy-Storage Applications. *ACS Appl Nano Mater*. **1**, 851–859 (2018).

51.    Q. Ji, C. Su, N. Mao, X. Tian, J. C. Idrobo, J. Miao, W. A. Tisdale, A. Zettl, J. Li, J. Kong, Revealing the Brønsted-Evans-Polanyi relation in halide-activated fast MoS2 growth toward millimeter-sized 2D crystals. *Sci Adv*. **7** (2021).

52.    A. A. Murthy, P. Masih Das, S. M. Ribet, C. Kopas, J. Lee, M. J. Reagor, L. Zhou, M. J. Kramer, M. C. Hersam, M. Checchin, A. Grassellino, R. Dos Reis, V. P. Dravid, A. Romanenko, Developing a Chemical and Structural Understanding of the Surface Oxide in a Niobium Superconducting Qubit. *ACS Nano*. **16**, 17257–17262 (2022).

53.    Georg. Brauer, *Handbook of Preparative Inorganic Chemistry V2.* (Elsevier Science, ed. 2nd, 1965).

54.    P. Yang, S. Zhang, S. Pan, B. Tang, Y. Liang, X. Zhao, Z. Zhang, J. Shi, Y. Huan, Y. Shi, S. J. Pennycook, Z. Ren, G. Zhang, Q. Chen, X. Zou, Z. Liu, Y. Zhang, Epitaxial Growth of Centimeter-Scale Single-Crystal MoS2 Monolayer on Au(111). *ACS Nano*. **14**, 5036–5045 (2020).

55.    C. Li, T. Kameyama, T. Takahashi, T. Kaneko, T. Kato, Nucleation dynamics of single crystal WS2 from droplet precursors uncovered by in-situ monitoring. *Scientific Reports 2019 9:1*. **9**, 1–7 (2019).

56.    A. Rothman, V. G. Dubrovskii, E. Joselevich, Kinetics and mechanism of planar nanowire growth. *Proc Natl Acad Sci U S A*. **117**, 152–160 (2020).

57.    F. Panciera, Y. C. Chou, M. C. Reuter, D. Zakharov, E. A. Stach, S. Hofmann, F. M. Ross, Synthesis of nanostructures in nanowires using sequential catalyst reactions. *Nature Materials 2014 14:8*. **14**, 820–825 (2015).

58.    B. J. Kim, J. Tersoff, S. Kodambaka, M. C. Reuter, E. A. Stach, F. M. Ross, Kinetics of individual nucleation events observed in nanoscale vapor-liquid-solid growth. *Science (1979)*. **322**, 1070–1073 (2008).

59.    E. Bakkers, Bottom-up grown nanowire quantum devices. *MRS Bull*. **44**, 403–410 (2019).

**Supplemental Information**

**Salt-assisted vapor-liquid-solid growth of one-dimensional van der Waals materials**

Thang Pham[1], Kate Reidy[1], Joachim D. Thomsen[1], Baoming Wang[1], Nishant Deshmukh[2], Michael A. Filler[2] and Frances M. Ross[1]

1. Department of Materials Science and Engineering, Massachusetts Institute of Technology, Cambridge, MA 02139, USA
2. School of Chemical & Biomolecular Engineering, Georgia Institute of Technology, Atlanta, GA 30332, USA

Corresponding authors. Email: thangpt88@gmail.com (T.P.), michael.filler@chbe.gatech.edu (M.A.F.), fmross@mit.edu (F.M.R.)

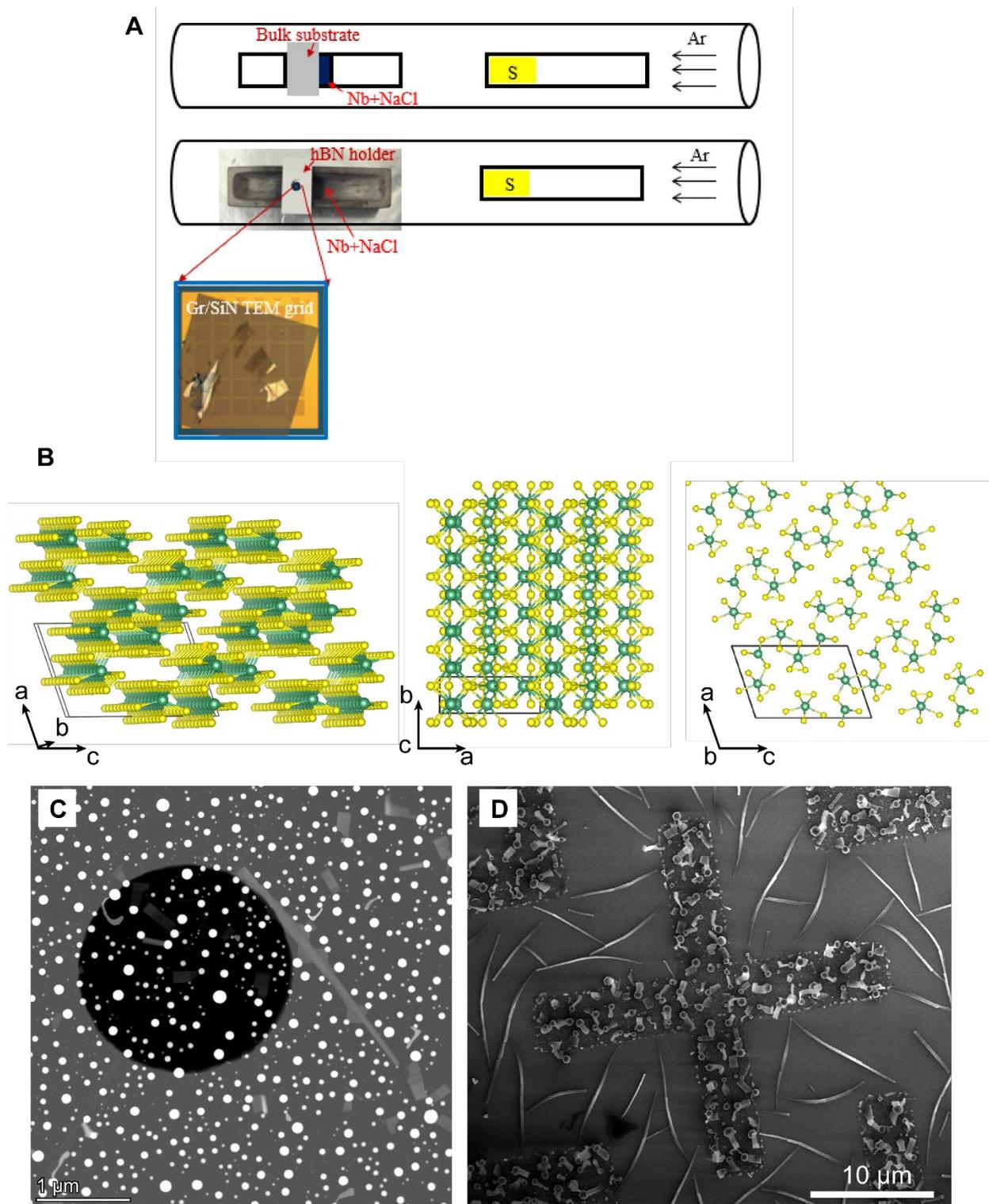

**Fig. S1. Overview of VLS growth of NbS₃ 1D nanostructures.** (A) Setup of the growth using a horizontal tube furnace, for growths on both bulk substrates (top) and on vdW materials, in this case few-layer graphene, suspended on a SiN TEM grid (bottom). (B) Atomic model of NbS₃ viewing along different crystallographic orientation. (C) Low-magnification STEM image of 30-min growth on suspended graphene. (D) Patterned growth using 50-nm Au patterned

lithographically on a SiO$_x$/Si substrate. Note the low density of uncatalyzed NWs growing in regions without Au.

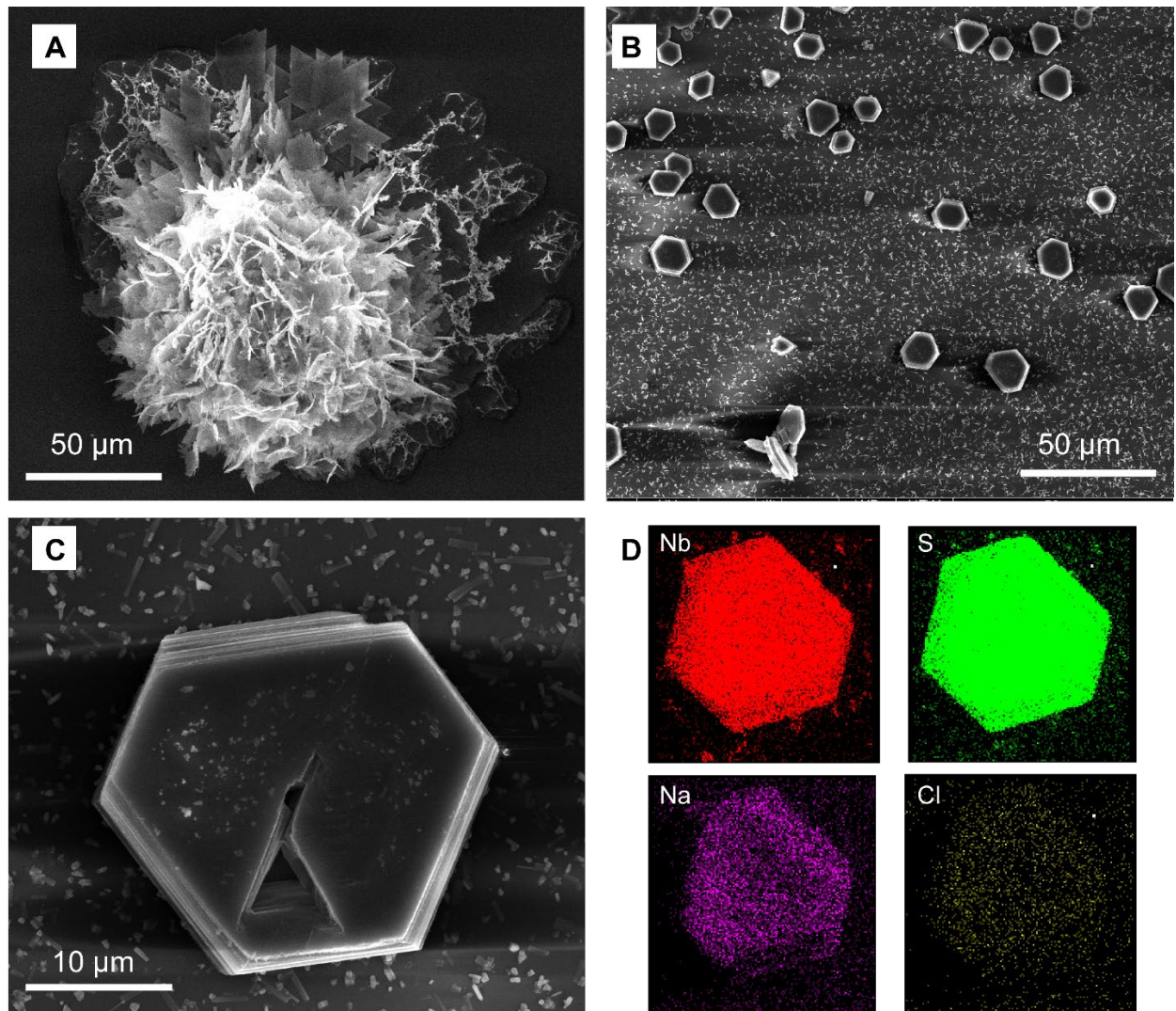

**Fig. S2. NbS$_2$ phase formed during high-temperature growth.** NbS$_2$ phase seen in the form of nano-flowers (A) and hexagonal plates (B) after growths at temperature above 750°C. The amount of precursors and other growth parameters were the same as used elsewhere. (C, D) Higher magnification SEM image of a hexagonal-plate multilayer NbS$_2$ and its corresponding SEM-EDS elemental mapping showing the presence of salt in the grown structure. These morphologies are characteristic of NbS$_2$ with its hexagonal crystal structure and are not shown by NbS$_3$.

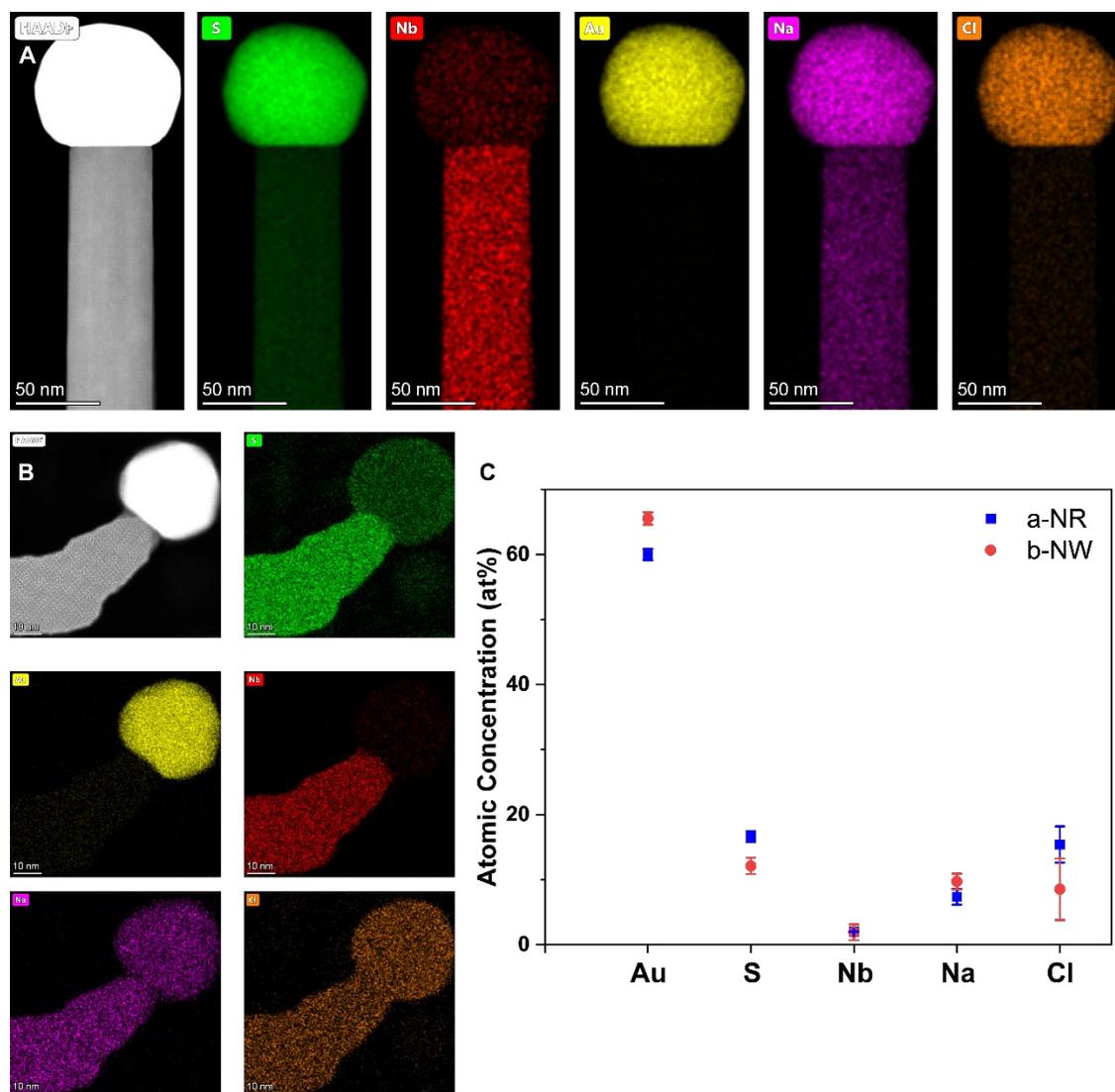

**Fig. S3. Chemical compositions of salt-assisted VLS NbS₃ structures**. STEM-EDS elemental mapping of a (A) b-NW and (B) a-NR. (C) Atomic concentration of different elements, including Au, S, Nb, Na and Cl, in the post-growth catalyst particles of a-NR and b-NW.

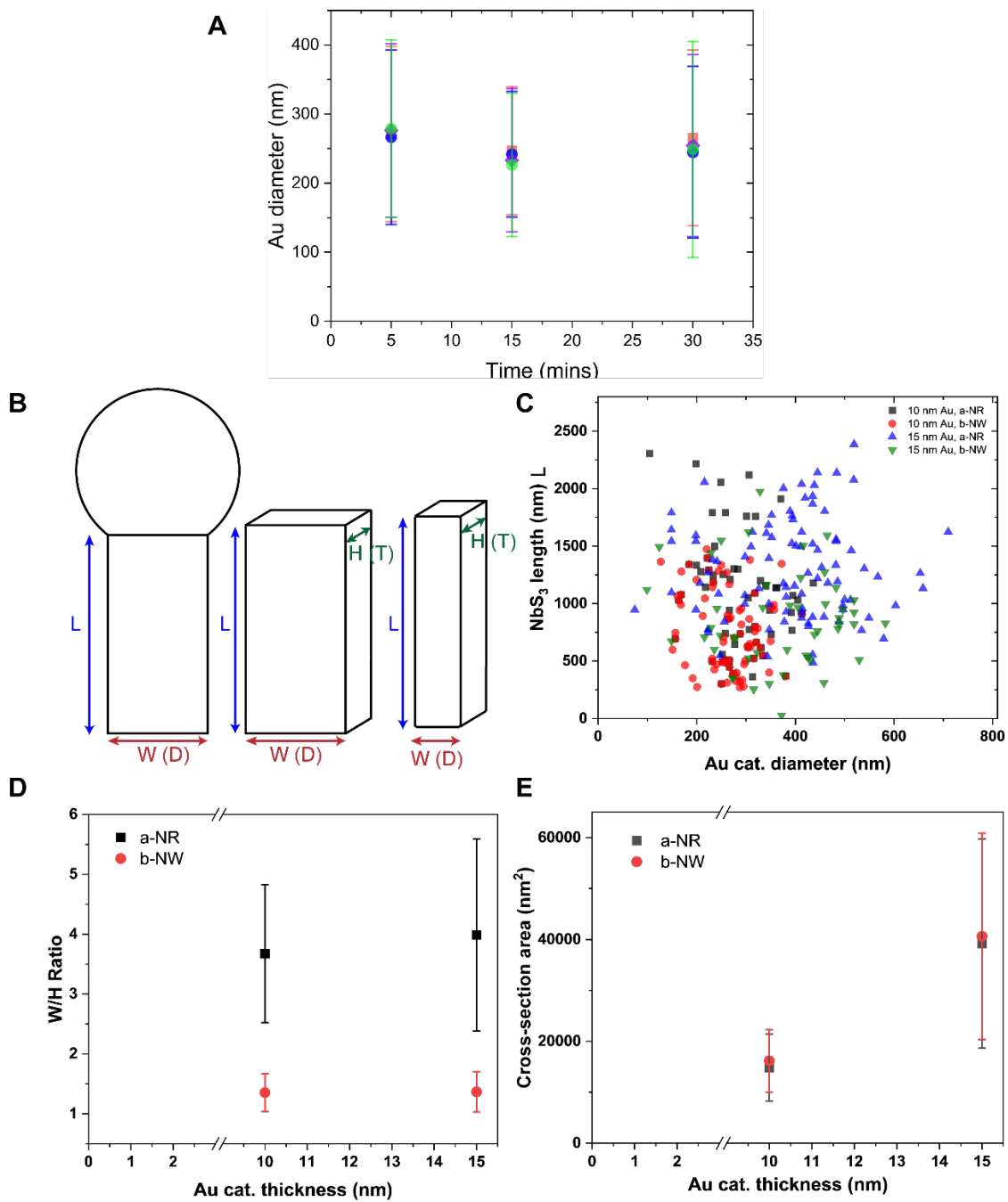

**Fig S4. Measured dimensions of Au catalysts, a-NRs and b-NWs.** (A) Au catalyst diameters at different growth times (5, 15 and 30 mins). The other growth parameters were kept the same. (B) Schematics showing the geometries of a-NB and b-NW and our definitions of length (L), width or diameter (W or D) and height or thickness (H or T) used in this study. Two growths using 10-nm and 15-nm Au films were used to obtain good statistics on the measurement of (C) NbS$_3$ length, (D) Width/Height (W/H) ratio and (E) cross-section area of the 1D materials. For W/H in (B), width is defined as the longer side in the cross section and height as the shorter side. The cross section area in (E) was measured at the visible end of the 1D materials.

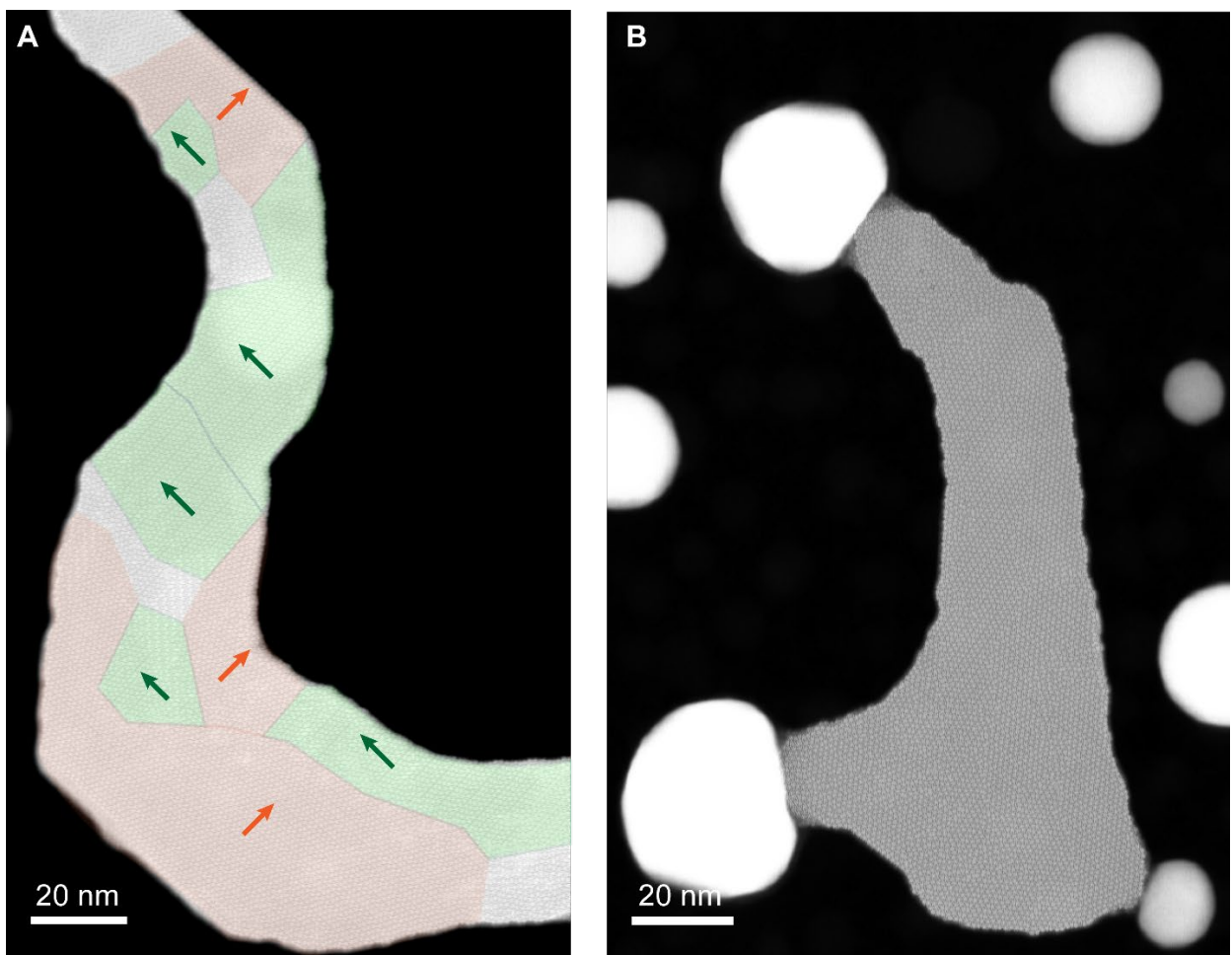

**Fig. S5. Stacking disorder in a-NRs.** (A) Colored polygons represent domains with different bilayer stacking orientation, which is defined and illustrated by the arrows. There is a rotation and rigid body shift between adjacent domains, for example a rotation between orange and green domains at the bottom left corner, and a shift between two adjacent green domains in the middle of the image. In uncolored regions, there is disorder between chain-chain interaction and no clear bilayer structure. (B) Another example of a nanoribbon having stacking disorder.

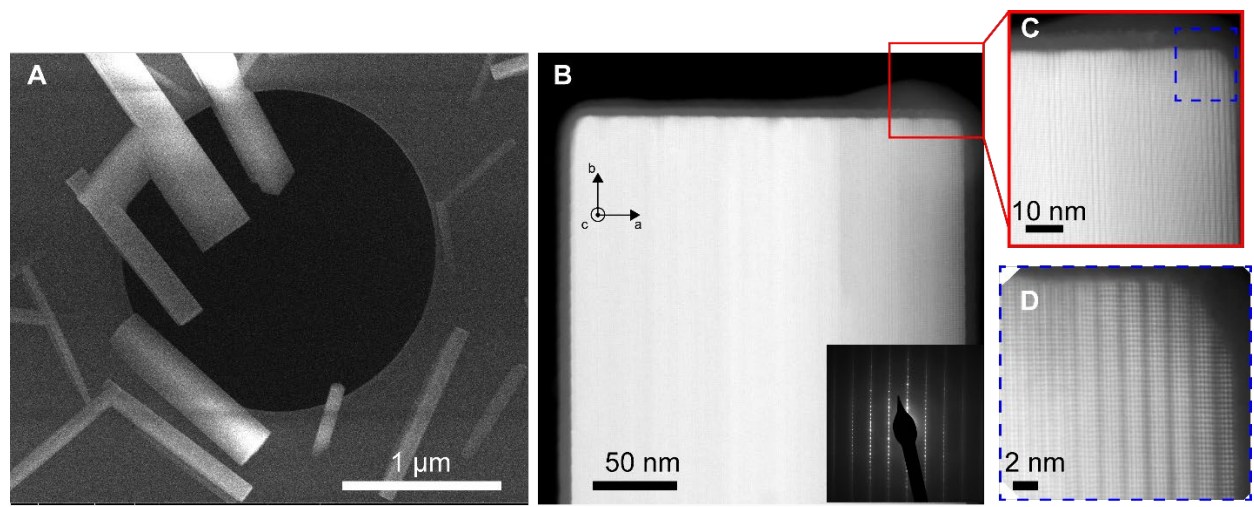

**Fig. S6. 1D NbS₃ synthesized in the absence of Au catalyst.** (A) STEM image of NbS₃ 1D nanomaterials grown directly on a SiN TEM grid. Note that the morphology is similar to the materials grown on SiOx/Si substrate (Fig. 4A) using the same growth conditions. (B) High-resolution STEM image of a representative structure showing its growth orientation of [010] (b-NW). (Inset) Its corresponding electron diffraction. (C, D) Atomic-resolution STEM images at increasing magnification showing the chain structure of the material.

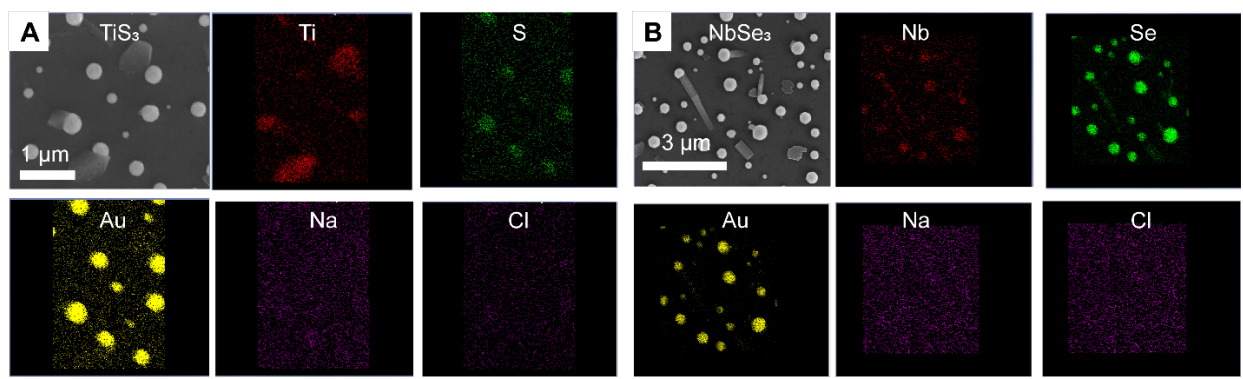

**Fig. S7. Chemical analysis of other TMT 1D nanostructures**. SEM-EDS chemical mapping of (A) TiS₃ and (B) NbSe₃ after salt-assisted VLS synthesis. The low signal/noise in SEM-EDS results in some of the chemical maps being noisy. For TiS₃ growth in (A), the growth temperature is 725°C and 50 mg of TiO₂ was used the metal precursor. The other synthesis parameters were kept the same as in the growth of NbS₃. For NbSe₃ growth in (B), the growth temperature is 825°C.

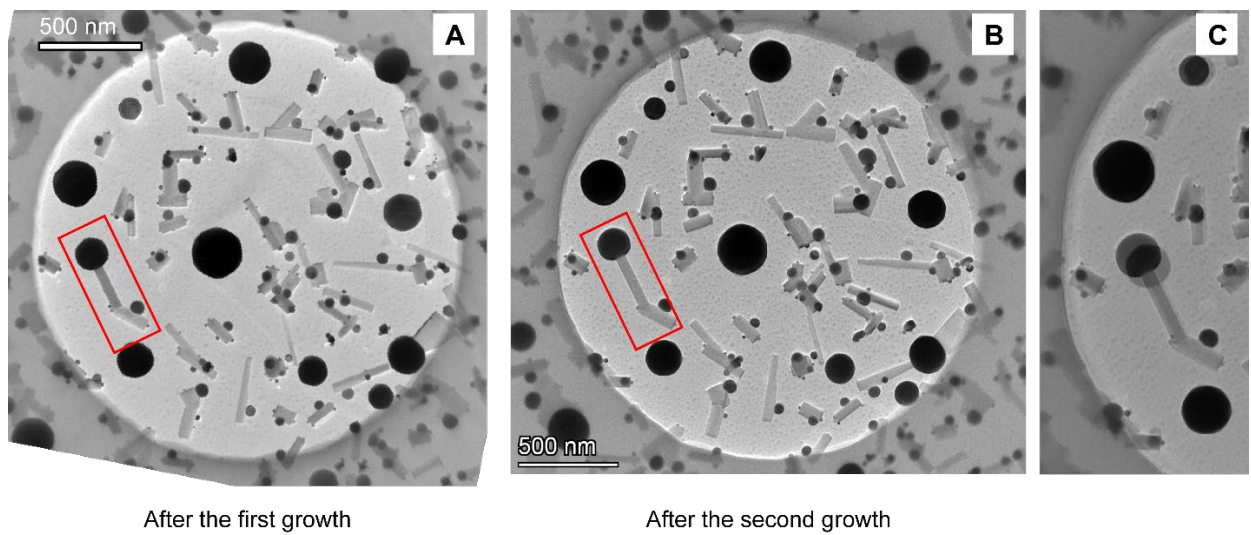

After the first growth          After the second growth

**Fig S8. Interrupted growth of b-NW NbS₃.** (A) TEM image of NbS$_3$ nanostructures grown on graphene transferred onto a SiN TEM grid at 725°C for 20 mins, using 3-nm Au catalyst film. (B) The same area imaged after a second growth for 20 mins using the same synthesis conditions (and no additional Au deposited). The b-NW enclosed by the red rectangle grew longer after the second growth. Note that its catalyst shape also changes, and other small motions of Au are visible elsewhere. (C) The overlapped images show that the b-NW grew from the catalyst-nanowire interface (the catalyst particle moved closer to the SiN grid's circular edge), adding 68 nm in length over the 20 mins of the second growth.